\documentclass[10pt]{article}
\newcommand{\bef}{\begin{figure}}
\newcommand{\eef}{\end{figure}}
\usepackage{aas_macros} 
\usepackage{astrobib} 

\usepackage{graphicx}

\begin{document}

\title{Neutron Stars: Laboratories for fundamental physics under extreme 
astrophysical conditions}


\author{Debades Bandyopadhyay$^{1,2}$\\
$^1$Astroparticle Physics and Cosmology Division, \\Saha Institute of Nuclear 
Physics,\\ 1/AF Bidhannagar, Kolkata-700064\\
$^2$ Homi Bhabha National Institute, Training School Complex,\\
Anushaktinagar, Mumbai-400094\\
{\em email: debades.bandyopadhyay@saha.ac.in}
}



\maketitle



\begin{abstract}
We discuss different exotic phases and components of matter from the crust to 
the core of neutron stars based on theoretical models for equations of state 
relevant to core collapse supernova simulations and neutron star 
merger. Parameters of the models are constrained from laboratory experiments. 
It is observed that equations of state involving strangeness degrees of 
freedom such as hyperons and Bose-Einstein condensates are compatible with 
2M$_{solar}$ neutron stars. The role of hyperons is explored on the evolution 
and stability of the protoneutron star (PNS) in the context of SN1987A. Moment 
of inertia, mass and
radius which are direct probes of neutron star interior are computed and their 
observational consequences are discussed. We continue our study on the dense 
matter under strong magnetic fields and its application to magnetoelastic 
oscillations of neutron stars.  
\end{abstract}

{\bf keywords:} neutron stars: eos, magnetic fields




\section{Introduction}
James Chadwick wrote to Niels Bohr about the discovery of the neutron 
in a letter dated 24 February, 1932 \cite{yakov}. The paper on the discovery 
of the neutron was published in {\it Nature} on 27 February, 1932. It is 
amazing to note that Lev Landau 
thought of a highly dense astrophysical object as a giant nucleus in 1931 well 
before this discovery and wrote an article on this subject which was published 
almost at the same time of the discovery of the neutron on 29 February, 1932 
\cite{land32}. 
In the Stanford meeting of the American Physical Society 
in 1933, Baade and Zwicky declared \cite{bz34} "With all reserve we advance the 
view that supernovae 
represent the transition from ordinary stars to neutron stars which in their
final stages consist of extremely closely packed neutrons." These developments
marked the beginning of research in physics and astrophysics of
neutron stars \cite{yakov}. 

Shortly after the discovery of a pulsar in 1967 \cite{hewis68}, the study of 
dense matter in the core of neutron stars gained momentum.
With the advent of x-ray, gamma-ray and radio telescopes, the
observational study of neutron stars has entered into a
new era. Space based Indian observatory ASTROSAT is the newest addition in
this pool. Observations using these facilities as well as other
telescopes are pouring in very exciting data on neutron stars.
From those observations, it might be possible to estimate masses,
radii, moment of inertia, surface temperatures and magnetic fields of neutron 
stars \cite{konar16}. Next generation radio
telescope known as Square Kilometer Array (SKA) to be co-located in
South Africa and Australia.
With the detection of gravitational wave signal from the event in GW150914 by
LIGO observatory, gravitational wave astrophysics opens a new window to probe
the neutron star interior. 
It would be possible to study fundamental physics in strong gravitational 
fields of pulsars and black holes using the SKA and LIGO-India along with other
telescopes.

Neutron stars harbour the densest form of matter in its interior. 
These compact astrophysical objects are unique laboratories for cold and dense 
matter. For such a cold
and dense matter can not be produced in terrestrial laboratories.
Wide range of density, from the density of iron nucleus at the surface of the 
star to several times normal nuclear matter density 
(2.7 $\times 10^{14}$ g/cm$^3$) in the core are expected to be present in 
neutron stars. 
The composition and structure of a neutron star are determined by the nature of 
strong interaction. 
Several novel phases with large strangeness fraction such as, hyperon
matter \cite{glen92,glen96,dc}, Bose-Einstein condensates of strange mesons
\cite{kap,pal2,Bani,knor} and quark matter \cite{farhi} may appear in the high 
density 
regime in neutron stars due to Pauli exclusion principle. Furthermore,
the recent accurately measured 2.01$\pm 0.04$ M$_{solar}$ neutron star puts 
stringent condition on the composition and equation of state (EoS) \cite{anto}. 

On the other hand, there is a growing interplay between the physics of dense
matter found in laboratories and neutron stars.
Though the Quantum Chromodynamics (QCD) predicts a very rich phase structure of
dense matter, we can only probe a small region of it in laboratories.
Relativistic heavy ion experiments produce a
hot (a few hundreds MeV) and dense matter (a few times normal nuclear matter
density). The study of dense matter in heavy ion collisions reveals many new
and interesting results such as the modifications of hadron
properties in dense
medium, the properties of strange matter including hyperons and (anti)kaons and
the formation of quark-gluon plasma \cite{rmp16,rmp17}. These empirical 
information from heavy ion collisions may be useful in understanding dense 
matter in neutron star interior. Properties of finite nuclei obtained in 
laboratories such as incompressibility of matter, symmetry energy etc also 
contribute to the understanding of matter in neutron stars.
  
Extremely high magnetic fields might be produced in heavy ion collisions when
moving charges of two heavy nuclei say Gold or Lead collide with each other at 
the speed of light. It was estimated that this field could be as high as 
$10^{19}$ G \cite{khar}. However, such
a strong magnetic field is produced for a short time $\sim$ a few fm/c. On the 
other hand, it was observed that a new class of neutron stars known as 
magnetars had very strong surface magnetic fields $\sim 10^{15}$ G. It was
inferred from the virial theorem \cite{chandra53} that the interior 
magnetic field could be several times higher than the surface fields of 
magnetars.

This shows that neutron stars are unique laboratories for fundamental physics
under extreme densities, magnetic fields and strong gravitational fields. In 
this article, we describe different phases of matter in supernova 
simulations and neutron stars and discuss how compositions and EoS of matter
can be constrained from observations. In Section 2, theoretical models of EoS
in the crust and core are introduced. In connection to SN1987A, the application
of this EoS in supernova simulations is elaborated in Section 3. Calculations 
of mass, radius and moment of inertia and their observable consequences are
presented in Section 4. Matter in strong magnetic fields and oscillatory 
modes of magnetars are discussed in Section 5. Finally conclusions are drawn in
Section 6.        

\section{Theoretical modeling of EoS}

\subsection{Matter in Neutron Star Crust}

Neutron star interior is broadly separated into two regions - crust and core. 
Again the  
crust is divided into the outer and inner crust; so is the core. There is a 
huge variation of matter density starting from $10^{4}$ g/cm$^3$ in the outer 
crust to
$\sim 10^{15}$ g/cm$^3$ in the core. Consequently, this leads to interesting
phases and compositions of matter in different layers of neutron stars. The 
outer crust is composed of nuclei in the background of a uniformly distributed
relativistic electron gas. Around $4 \times 10^{11}$ g/cm$^3$, neutrons start 
dripping out of nuclei when the neutron
chemical potential is equal to bare neutron mass. This is the end of the
outer crust and beginning of the inner crust. In this layer of matter, 
the components of matter are neutron-rich nuclear cluster, free neutrons and
a uniform gas of relativistic electron gas. As the density increases, the matter
passes through an interesting phase called the pasta phase where various 
geometrical shapes such as rod, slab, bubble etc might appear due to 
competition between the surface tension and Coulomb interaction in nuclear 
clusters. It shows that the matter is highly non-uniform in neutron star 
crusts. Neutron-rich nuclear clusters dissolve into neutrons and 
protons which ,in turn, produce a uniform nuclear matter, at the crust-core
interface around the matter density $2.7 \times 10^{14}$ g/cm$^3$.  
  
We introduce here the nuclear statistical equilibrium (NSE) model for the
description of matter of light and heavy nuclei together with unbound but
interacting nucleons at low temperature and mass density below 
$\sim 2.7 \times 10^{14}$ g/cm$^3$
\cite{hs}. In this model, the nuclear chemical equilibrium
is regulated by the modified the Saha equation.       
The total canonical partition function in this model is given by
\begin{equation}
Z(T,V,\{N_i\})=Z_{\rm nuc}~\prod_{A,Z}Z_{A,Z}~Z_{\rm Coul} \; ,
\end{equation}
with $V$ denoting the volume of the system.
The Helmholtz free energy involving free energies of nucleons ($F_{\rm nuc}$), 
nuclei ($F_{A,Z}$) and Coulomb ($F_{\rm Coul}$) is computed as,
\begin{eqnarray}
F(T,V,\{N_i\})&=&-T \mathrm{ln} Z \\
&=& F_{\rm nuc}+\sum_{A,Z} F_{A,Z} +F_{\rm Coul} \; .
\end{eqnarray}
The number density of each nuclear species (A,Z) is obtained from modified 
Saha equation \cite{banik14} 
\begin{eqnarray}
&&n_{A,Z}=\kappa~g_{A,Z}(T) 
\left(\frac{M_{A,Z} T}{2\pi}\right)^{3/2}\exp\left(\frac{(A-Z)\mu_{n}^0+Z\mu_{p}^0-M_{A,Z}-E^{\rm Coul}_{A,Z}-P^0_{\rm nuc}V_{A,Z}}T\right) \; , \label{eq_naz}
\end{eqnarray}
where $g_{A,Z}$ is the nuclear spin degeneracy; $\kappa$ is the volume fraction 
available for nuclei and approaches to
zero at the crust-core boundary. Finally one obtains the 
energy density and pressure in this model.

\subsection{Dense Matter in Neutron Star Core}

Neutrons and protons in neutron star core become relativistic as baryon density
increases. Furthermore, dense matter in neutron star interior is a highly 
many body system. The QCD might be the fundamental theory to describe such a 
dense matter. Here we focus on a relativistic field theoretical model involving 
baryons and mesons. In this Lorentz covariant theory, baryon-baryon interaction
is mediated by the exchanges of mesons. Meson-baryon couplings are made density
dependent. Being a relativistic model, this ensures causality in the 
EoS. 

The starting point in the density dependent relativistic hadron (DDRH) field 
theory is the Lagrangian density which describes baryon-baryon interaction 
through exchanges of scalar $\sigma$, vector $\omega$, $\phi$ and $\rho$ 
mesons \cite{banik14,typ10}, 

\begin{eqnarray}
\label{eq_lag_b}
{\cal L}_B &=& \sum_B \bar\psi_{B}\left(i\gamma_\mu{\partial^\mu} - m_B
+ g_{\sigma B} \sigma - g_{\omega B} \gamma_\mu \omega^\mu 
- g_{\phi B} \gamma_\mu \phi^\mu 
-  g_{\rho B} 
\gamma_\mu{\mbox{\boldmath $\tau$}}_B \cdot 
{\mbox{\boldmath $\rho$}}^\mu  \right)\psi_B\nonumber\\
&& + \frac{1}{2}\left( \partial_\mu \sigma\partial^\mu \sigma
- m_\sigma^2 \sigma^2\right)
-\frac{1}{4} \omega_{\mu\nu}\omega^{\mu\nu}\nonumber\\
&&+\frac{1}{2}m_\omega^2 \omega_\mu \omega^\mu
-\frac{1}{4} \phi_{\mu\nu}\phi^{\mu\nu}
+\frac{1}{2}m_\phi^2 \phi_\mu \phi^\mu \nonumber\\
&&- \frac{1}{4}{\mbox {\boldmath $\rho$}}_{\mu\nu} \cdot
{\mbox {\boldmath $\rho$}}^{\mu\nu}
+ \frac{1}{2}m_\rho^2 {\mbox {\boldmath $\rho$}}_\mu \cdot
{\mbox {\boldmath $\rho$}}^\mu.
\end{eqnarray}
Here $\psi_B$ denotes the baryon octets, ${\mbox{\boldmath 
$\tau_{B}$}}$ is the isospin operator and $g$s are density dependent 
meson-baryon couplings. It is to be noted that $\phi$ mesons are mediated 
between particles having strangeness quantum number. 

Next we can calculate the grand-canonical thermodynamic potential per unit 
volume 
\begin{eqnarray}
\frac{\Omega}{V} &=& \frac{1}{2}m_\sigma^2 \sigma^2
- \frac{1}{2} m_\omega^2 \omega_0^2 
- \frac{1}{2} m_\rho^2 \rho_{03}^2  
- \frac{1}{2} m_\phi^2 \phi_0^2 
- \Sigma^r \sum_{B} n_B
\nonumber \\
&& - 2T \sum_{i=n,p,\Lambda,\Sigma^-,\Sigma^0, \Sigma^+,\Xi^-,\Xi^0} \int \frac{d^3 k}{(2\pi)^3} 
[\mathrm{ln}(1 + e^{-\beta(E^* - \nu_i)}) +
\mathrm{ln}(1 + e^{-\beta(E^* + \nu_i)})] ~,  
\end{eqnarray}
where the temperature is defined as $\beta = 1/T$ and
$E^* = \sqrt{(k^2 + m_i^{*2})}$.
This involves a term called
the rearrangement term $\Sigma^{r}$ \cite{banik14,hof} due to many-body
correlations which is given by
\begin{equation}\label{eq_rear}
\Sigma^{r}=\sum_B[-g_{\sigma B}'  
\sigma n^{s}_B + g_{\omega B}' \omega_0 
n_B
+ g_{\phi B}' \phi_0 
n_B
+ g_{\rho B}'\tau_{3B} \rho_{03} n_B
+   g_{\phi  B}' \phi_0 n_B  ]~,
\end{equation}
where $'$ denotes derivative with respect to baryon density of species B.

We also study the Bose-Einstein condensation of antikaons ($K^-$ mesosn) in
neutron star matter. In this case, baryons are embedded in the condensate.
We treat the kaon-baryon interaction in the same footing as the baryon-baryon 
interaction described by the Lagrangian density (\ref{eq_lag_b}).    
The Lagrangian density for (anti)kaons in the
minimal coupling scheme is \cite{glen99,bani01},
\begin{equation}
{\cal L}_K = D^*_\mu{\bar K} D^\mu K - m_K^{* 2} {\bar K} K ~,
\end{equation}
where $K$ and $\bar K$ denote kaon and (anti)kaon doublets; the covariant 
derivative is
$D_\mu = \partial_\mu + ig_{\omega K}{\omega_\mu} 
+ ig_{\phi K}{\phi_\mu} 
+ i g_{\rho K}
{\mbox{\boldmath t}}_K \cdot {\mbox{\boldmath $\rho$}}_\mu$ and
the effective mass of (anti)kaons is
$m_K^* = m_K - g_{\sigma K} \sigma$.
The thermodynamic potential for antikaons is given by,
\begin{equation}
\frac {\Omega_K}{V} = T \int \frac{d^3p}{(2\pi)^3} [ ln(1 - 
e^{-\beta(\omega_{K^-} - \mu)}) + 
 ln(1 - e^{-\beta(\omega_{K^+} + \mu)})]~.
\end{equation}
The in-medium energy of $K^{-}$ meson is given by
\begin{equation}
\omega_{K^{-}} =  \sqrt {(p^2 + m_K^{*2})} - \left( g_{\omega K} \omega_0
+ g_{\phi K} \phi_0
+ \frac {1}{2} g_{\rho K} \rho_{03} \right)~,
\end{equation}
and $\mu$ is the chemical potential of $K^-$ mesons and is given by
$\mu = \mu_n -\mu_p = \mu_e$. The threshold condition for s-wave ({\bf p} = 0)
$K^-$ condensation is given
by
$\mu = \omega_{K^{-}} =   m_K^* - g_{\omega K} \omega_0 - g_{\phi K} \phi_0
- \frac {1}{2} g_{\rho K} \rho_{03}$~. Mean field values of mesons are $\sigma$,
$\omega_0$, $\phi_0$ and $\rho_{03}$. 

Thermodynamic quantities like energy density, pressure etc in the hadronic 
and kaon condensed phases are computed from the grand-thermodynamic potentials
\cite{banik14,char14,bani08}. Charge neutrality and $\beta$-equilibrium 
constraints are imposed on neutron star matter. 

Finally, meson-nucleon density dependent couplings are obtained by fitting 
properties of finite nuclei \cite{banik14,typ10}. Vector meson 
couplings for hyperons and kaons are estimated theoretically using the symmetry
relations \cite{weis,sch} whereas their scalar couplings are obtained from 
hyeprnuclei and kaonic atom data \cite{char14}.   

Recently, Banik, Hempel and 
Bandyopadhyay (BHB) constructed a hyperon EoS for supernova and neutron star
matter involving $\Lambda$ hyperons and the repulsive $\Lambda$-$\Lambda$
interaction mediated by $\phi$ mesons (Banik, Hempel \& Bandyopadhyay). This 
hyperon EoS is compatible with 2M$_{solar}$ neutron stars and denoted by
BHB$\Lambda \phi$ \cite{banik14}.   

In the following sections, we describe the role of compositions and EoS on the 
evolution of the PNS in core collapse supernova 
simulations, masses, radii and moments of inertia of neutron stars and 
magnetoelastic oscillations of strongly magnetised neutron stars.  
\begin{figure}
\begin{center}
\includegraphics[width=8cm,height=8cm]{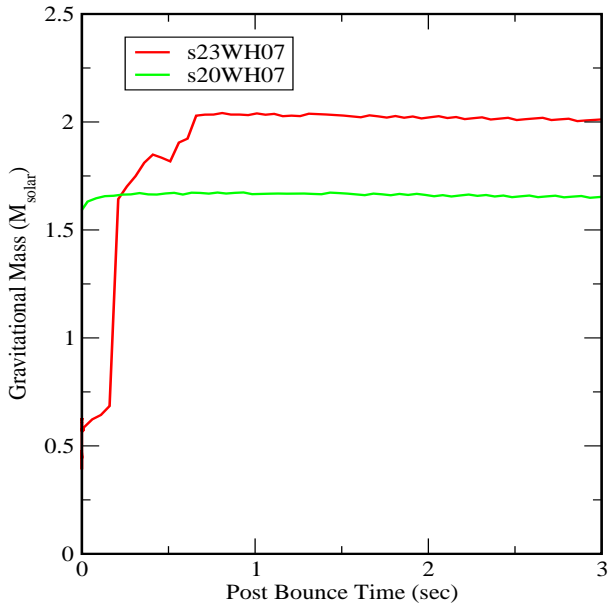}
\caption{Long duration evolution of the protoneutron star using 20 and 23 
M$_{solar}$ progenitors and BHB$\Lambda \phi$ EoS.}\label{figOne}
\end{center}
\end{figure}
\section{Mystery of the missing compact star in SN1987A}

Over the past thirty years, SN1987A has been the most studied core-collapse 
supernova event. It is the only supernova event from which neutrinos were 
detected after the explosion over 11 seconds. It was evident from the 
detection of neutrinos that a hot and neutrino-trapped protoneutron star 
was born and existed for about 11 s. There is no detection of a neutron star in
SN1987A so far. It is believed that an event horizon was formed after 11 s and
the PNS collapse into a black hole. The question is what made the PNS 
metastable and drove it into a black hole. 
 
Different groups investigated the problem of stability of a PNS for short 
times. When a PNS is made up of nucleons and leptons, it has a slightly smaller
maximum mass than that of the neutron star. However, this situation
changes with the appearance of exotic matter such as hyperons or 
$K^-$ condensation in dense matter during the evolution of the PNS
\cite{bani14,bethe}. The PNS including hyperon and/or Bose-Einstein 
condensate has a higher maximum mass than that of a cold neutron star 
\cite{bethe,pra95,bani01} . 
Neutrino and thermal pressure could stabilize much larger maximum mass for a 
protoneutron star during the evolution.  However, the PNS might be unstable 
after deleptonization and cooling.

The role of $\Lambda$ hyperons on supernova explosion mechanism and the 
evolution of PNS has been studied using a general 
relativistic one dimensional core collapse supernova model \cite{ott11}. 
Earlier simulations were done with the hyperon EoS which was not compatible with
the two solar mass neutron star \cite{bani14}. 
Furthermore, the long duration evolution of the PNS with enhanced neutrino 
heating in the supernova simulation with 23 solar mass progenitor denoted as
s23WH07 is 
investigated to test the hypothesis of metastability in the PNS. 
The $\Lambda$
hyperon EoS of Banik, Hempel and Bandyopadhyay BHB$\Lambda \phi$ is used as 
microphysical input in this simulation. $\Lambda$ hyperons appear just after 
core bounce and its population became significant as the PNS evolves. 
This simulation leads to a successful supernova explosion and the 
PNS evolves to a stable neutron star of 2.0 M$_{solar}$ over 3 sec as
evident from Figure {\ref{figOne}}. This is compared with the result of our 
earlier CCSN simulation of 20 M$_{solar}$ progenitor denoted as s20WH07 that 
led to a stable 
neutron star of 1.6 M$_{solar}$ \cite{char15}. These findings are at
odds with the prediction about the collapse of the PNS into a black hole after 
deleptonization and cooling. 
\bef
\begin{center}
      \includegraphics[height=6cm,width=6.0cm]{wkm.eps}
\caption{In-medium energy of $K^-$ mesons ($\omega_{K^-}$) and electron 
chemical potential ($\mu_e$) as a function of normalised baryon density.}\label{figTwo}
\end{center}
\eef
\begin{figure}
\begin{center}
\includegraphics[height=8cm,width=8cm]{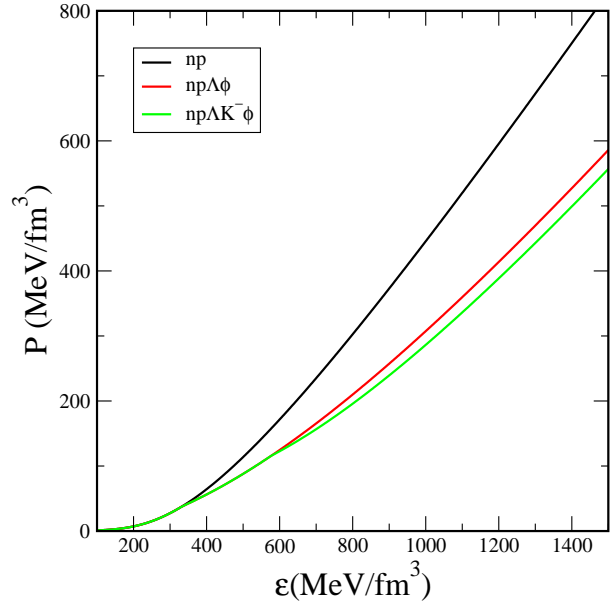}
\caption{Pressure as a function of energy density for compositions np, 
np$\Lambda \phi$ and np$\Lambda K^{-} \phi$.}\label{figThree}
\end{center}
\end{figure}
\section{Probing neutron star interior: Mass, Radius and Moment of Inertia}

Neutron star masses have been estimated to very high degree of
accuracy due to the measurement of post Keplerian parameters in relativistic
binary systems. The accurately measured highest neutron
star mass ($M$) is 2.01$\pm 0.04$ M$_{\odot}$ so far. However, the estimation 
of radius from observations is still problematic \cite{sudip17}.
The discovery of highly relativistic binary systems such as the double
pulsar system PSR J0737-3039 for which masses of both pulsars are known
accurately, opens up the possibility for the determination of moment of 
inertia ($I$) of pulsar $A$ which ,in turn, might overcome the uncertainties 
in the determination of radius ($R$). It is expected that
the high precision timing technique in the upcoming 
SKA would facilitate the extraction of the moment of
inertia of a pulsar earlier than that in the present day scenario.
Higher order post Newtonian (PN) effects in relativistic neutron star 
binaries could be probed in the SKA era. Furthermore, the relativistic 
spin-orbit (SO) coupling might result in an extra advancement of periastron 
above the PN contributions. The measurement of the SO coupling effect over and
above the contribution of the second PN term could lead to 
the determination of moment of inertia of a pulsar in 
relativistic neutron star binaries in general \cite{damo} and the double 
pulsar system in particular \cite{schutz}.
Observed masses, radii and moments of neutrons are direct probes of compositions
and EoS in neutron star interior. The theoretical mass-radius, moment of
inertia - compactness parameter (ratio of mass and radius) relationships of 
neutron stars could be directly compared with measured masses,
radii and moments of inertia from various observations. Observations indicate
that neutron stars are slowly rotating and the fastest rotating neutron star 
among them has a frequency 716 Hz. 
\begin{figure}
\begin{center}
\includegraphics[width=8cm,height=8cm]{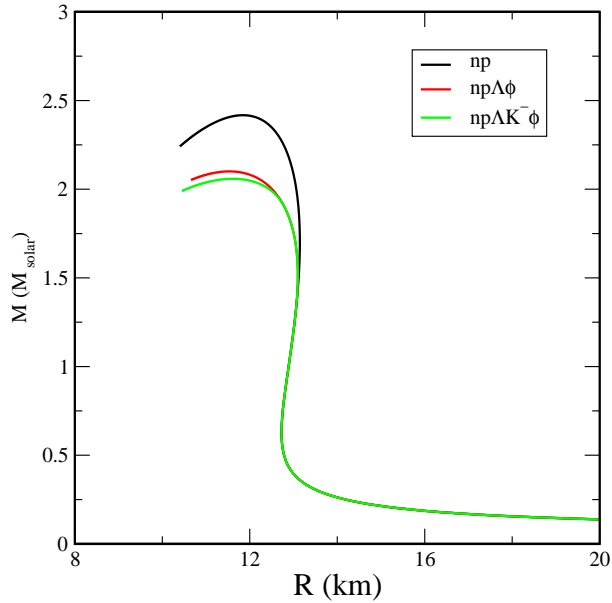}
\caption{Mass-Radius relationship for neutron star compositions np, 
np$\Lambda\phi$ and np$\Lambda K^{-}\phi$.}\label{figFour}
\end{center}
\end{figure}
Structures of non-rotating neutron stars are computed from the 
Tolman-Oppenheimer-Volkoff (TOV) equation,
\begin{eqnarray}
\frac{dp}{dr} &=& - {\frac{G \varepsilon (r) m(r)}{c^2 r^2}}
\left(1+\frac{p(r)}{\varepsilon(r)}\right) \left(1 + {\frac{4 \pi r^3 p(r)}{m(r) c^2}}\right)\nonumber\\ 
&& \times {\left[1 - {\frac{2 G m(r)}{c^2 r}}\right]}^{-1}~.
\end{eqnarray}
We need an EoS to close the TOV equation.

Slowly rotating neutron stars are investigated by perturbing the spherical 
space-time metric \cite{hart}. Moment of inertia is calculated from 
$I = J/{\Omega}$ where
\begin{equation}
I = {8 \pi \over 3} \int_0^R  r^4 e^{(\lambda - \nu)}
     \left( p (r)  +\varepsilon (r) \right) \frac{\left(\Omega - \omega (r) \right)}{\Omega}dr~,
\end{equation}

and the frame-dragging angular velocity ($\omega$) is obtained by solving the 
Hartle equation; $\Omega$ is the spin of the neutron star and $\lambda$, $\nu$
are metric functions.

We consider different compositions for the computation of EoS, mass-radius
relationship and moment of inertia. Neutron star matter made of neutrons and 
protons is denoted by $np$. In this calculation, $\Lambda$ hyperons appear 
first at baryon density $n_b = 2.2 n_0$ where the saturation density is 
$n_0 = 0.149 fm^{-3}$. The repulsive $\Lambda$-$\Lambda$ interaction is 
mediated by $\phi$ mesons. This composition of matter involving neutrons, 
protons and $\Lambda$ hyperons is represented by $np\Lambda\phi$.   
Being heavier, $\Sigma$ and $\Xi$ hyperons are populated at much higher 
densities and excluded from this calculation. Another exotic phase of matter
considered here is the Bose-Einstein condensed matter of $K^-$ mesons in which 
neutrons, protons and $\Lambda$ hyperons are embedded the condensate and 
denoted by 
$np\Lambda K^{-}\phi$. The threshold density for $K^-$ condensation is 
obtained from the equality of in-medium energy ($\omega_{K^-}$) of $K^-$ and 
electron chemical potential ($\mu_e$). This is exhibited in 
Figure {\ref{figTwo}}. In this case, the onset of the condensate occurs at
$n_b = 3.69 n_0$.     

\begin{figure}
\begin{center}
\includegraphics[width=8cm,height=8cm]{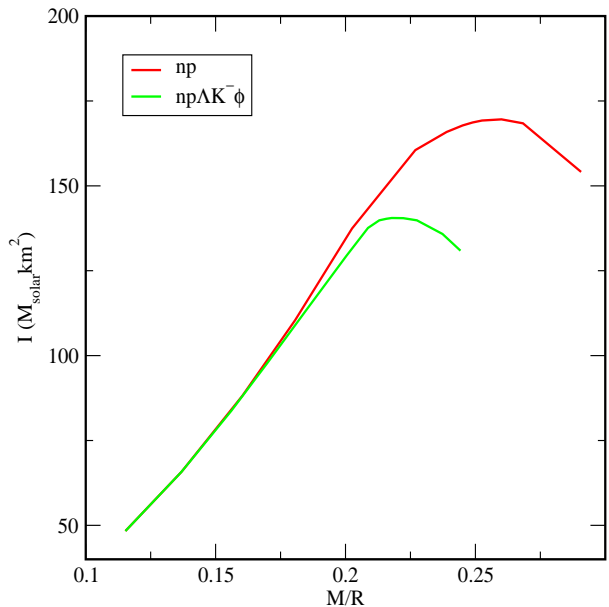}
\caption{Moment of inertia versus compactness for neutron star compositions np, 
and np$\Lambda K^{-}\phi$.}\label{figFive}
\end{center}
\end{figure}
Figure {\ref{figThree}} shows the relation between pressure ($P$) and energy
density ($\varepsilon$) which is known as the EoS, for the above mentioned
compositions of matter. It is evident from the figure that additional degrees
of freedom in the form of hyperons and $K^-$ condensate make an EoS softer.
This is also reflected in the structures of neutron stars. Mass-radius 
relationships for different compositions and EoS are shown in Figure 
{\ref{figFour}}. Being the stiffest among all other cases considered here,
nuclear matter EoS results in the highest maximum mass neutron star of 
2.42 M$_{solar}$. On the other hand, $\Lambda$ hyperons and $K^-$ condensate
make the EoS softer leading to small to smaller maximum mass neutron stars. The
maximum mass corresponding to $np\Lambda\phi$ case is 2.1 M$_{solar}$, whereas
it is 2.09 M$_{solar}$ for $np\Lambda K^{-}\phi$ case. It is important to note
that for exotic phases of matter maximum masses are well above the 
observational benchmark of 2.01$\pm 0.04$ M$_{solar}$. It demonstrates that
there is room for exotic matter in neutron star interior. Moment of inertia
is plotted against the compactness parameter ($M/R$) 
in Figure {\ref{figFive}}. It is evident from the figure that the moment of 
inertia
corresponding to nuclear matter EoS is significantly higher than that of the 
Bose-Einstein condensed matter for compactness above 0.2. 
If the moment of inertia of Pulsar A in the double pulsar
is estimated in future, the radius could be determined for this pulsar because 
its mass is already known accurately \cite{schutz}.    

\section{Neutron star matter in strong magnetic fields}

Neutron star crust plays an important role in many observational phenomena 
for example cooling of neutron stars, glitches and Quasi Periodic Oscillations
(QPOs). Heat 
transport and magnetic field evolution in the crust are sensitive to the 
composition of the crust. Similarly, the shear modulus which is an important 
input in understanding QPOs believed to magnetoelastic oscillations, is 
impacted by the crustal composition. On the other hand, superfluid neutrons in 
the crust might be responsible for pulsar glitches. 

It was observed that a class of neutron stars 
called magnetars had surface magnetic fields as large as $10^{15}$ G. Soft 
Gamma Repeaters (SGRs) and Anomalous X-ray Pulsars (AXPs) are thought to be 
very good candidates of magnetars \cite{duncan92,duncan98,kouv98}. SGRs 
exhibited giant flares of gamma rays in 
several instances. QPOs were observed in the decaying tails of giant flares in
SGR 0526-66, SGR 1900+14 and SGR 1806-20 caused by the magnetic field evolution
and its impact on the crust.

It was argued that the interior magnetic could be several orders of magnitude 
higher than the surface field of magnetars. The flux conservation in core
collapse supernovae and virial theorem \cite{chandra53} predict a maximum 
interior
magnetic field of $\sim 10^{18}$ G without causing any instability in the star.
Like a density gradient from the surface to the centre, the magnetic field 
should show a similar behaviour as described by the ansatz \cite{prl97b},
\begin{equation}
B_m(n_b/n_0) = B_m^{\rm surf} + B_0 \left[
1 - \exp\left\{ -\beta (n_b/n_0)^{\gamma} \right\} \right] ,
\end{equation}
\begin{figure}
\begin{center}
\includegraphics[height=8cm,width=8cm]{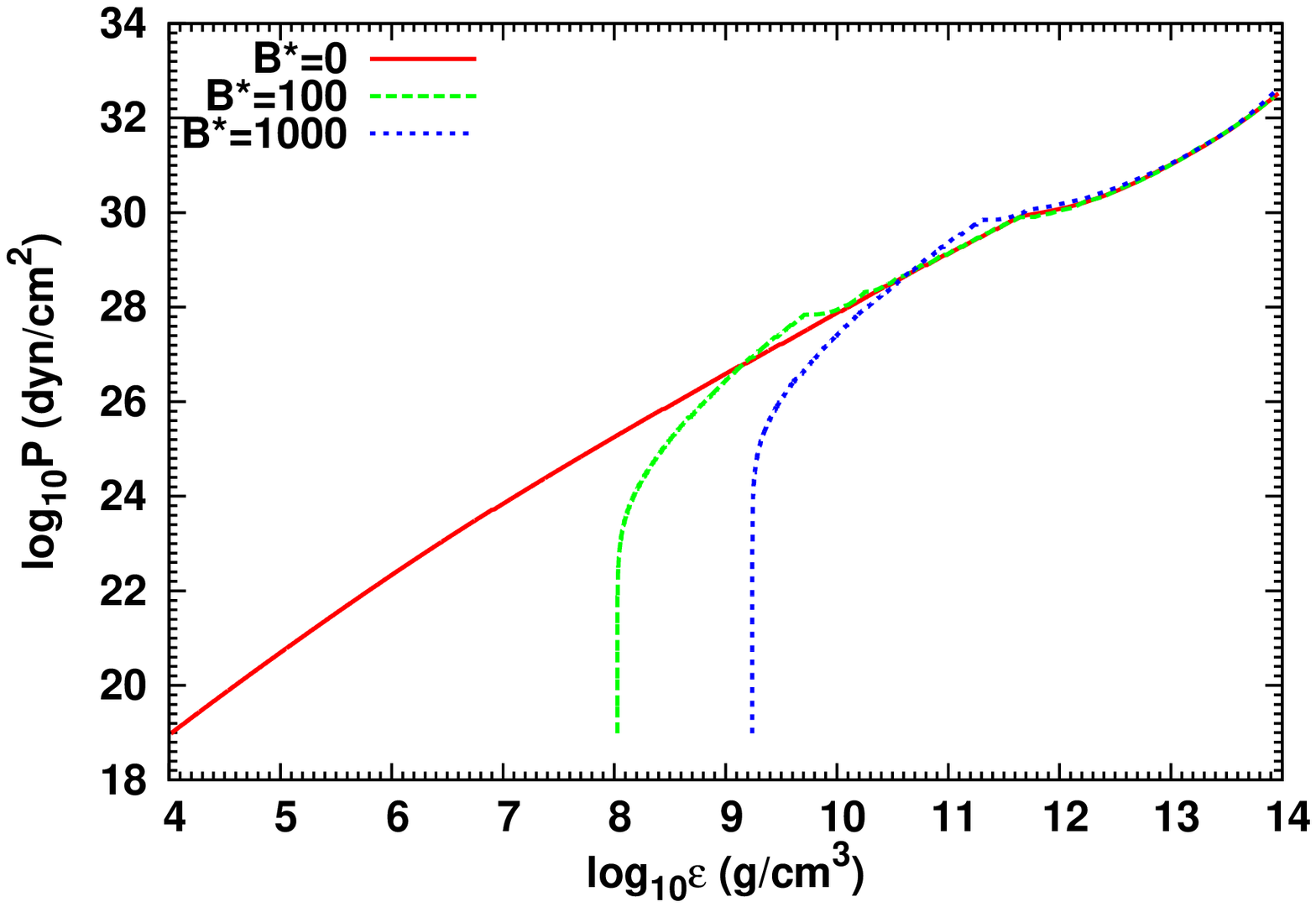}
\caption{EoS of neutron star crust with and without magnetic field.}\label{figSix}
\end{center}
\end{figure}
Several groups studied the influence of strong magnetic fields on the 
compositions and EoS of neutron star matter and its observable consequences
\cite{sc97,prl97b,db98,bro00,lai01,nandi16}. Such a strong 
magnetic field is expected to influence charged particles such as electrons in 
the crust through Landau quantization. As no free protons are available in the
crust, protons are not Landau quantised. However, protons are affected through
the charge neutrality. Number density, energy density and pressure of 
relativistic electrons are influenced by the phase space modifications of 
electrons due to Landau quantisation. Here we adopt the Baym, Pethick and
Sutherland (BPS) model of the outer crust \cite{nandi11a} and the inner crust 
model of Ref.(\cite{nandi11b}) in presence of strong magnetic fields. In 
Figure {\ref{figSix}}, pressure is plotted as a function of energy density for
the crust with and without magnetic fields. Here the magnetic field strength is 
given in terms of the critical field ($B_c$) for electrons i.e. $B = B_{*} B_c$
where $B_c = 4.414 \times 10^{13}$G. It is observed from the figure that the
EoS of the crust in presence of strong magnetic fields is significantly 
modified in the energy density regime $< 10^{10}$ g/cm$^3$ due to the 
population of electrons in the zeroth Landau level compared to the zero
field case ($B_{*}=0$). However, several Landau levels are populated in the
high density regime above $10^{10}$ g/cm$^3$. Consequently, results of 
$B_{*} =100, 1000$ approach the classical result without magnetic field. 
      
This magnetised crust model is applied to the problem of magnetoelastic 
oscillations in magnetars to explain QPOs in giant flares \cite{nandi16}. In
contrast to the state-of-the-art general relativistic magnetohydrodynamics
\cite{gabler}, our 
calculation is based on a general relativistic spherical symmetric model of
neutron stars with dipole magnetic fields and involves crust-core coupling.
Two situations are considered for magnetoelastic modes. In one case, 
magnetoelastic modes confined to the crust (CME) are relevant. In the other
case, global magnetoelastic (GME) modes become important when the crust-core
coupling is considered. For magnetic fields $> 10^{15}$G the Alfv\'en velocity 
becomes greater than the shear velocity. 
Consequently, GME mode frequencies just become those of pure Alfv\'en modes. 

Detections of fundamental and first overtone frequencies in SGR 1806-20
giant flare could constraint the EoS. This can lead to 
determination of the crust thickness. It was shown that the crust thickness
might be estimated from the ratio of fundamental and first overtone 
frequencies \cite{sotani} 
$\frac{\Delta R}{R} = {_{l}}C^n \frac{_{l}f^0}{_{l}f^n}$. 
It is also evident from this relation that the crust thickness is inversely 
proportional to the frequency of higher harmonics. One can estimate the crust
thickness taking 18 Hz as the fundamental frequency ($_{l}f^0$) and 626 Hz as
the first overtone frequency ($_{l}f^1$). This led to a ratio of 0.06
with $_{l}C^n \sim 2$  which favoured a stiff EoS model \cite{sotani}. It was 
noted that the radius of a neutron star increased in strong magnetic fields
compared with the zero field case. Consequently, the thickness of the crust
increased in strong fields \cite{nandi16}. We obtain a crust thickness of 
0.088 km and the value of $_{l}C^1$ is 3.06 for the magnetised EoS as shown in 
Fig. {\ref{figSix}}. Such a description relating the crust thickness to the 
ratio of observed frequencies is relevant for CME modes. The effects of 
magnetised crusts on magnetoelastic modes disappear above a critical field
$4 \times 10^{15}$ G. Furthermore, GME modes might explain all frequencies
of SGR 1806-20.  
  
\section{Conclusions and outlook}
We have demonstrated through core collapse supernova simulations and 
calculation of neutron star structures that EoSs involving exotic components of 
matter such as hyperons and/or Bose-Einstein condensates are compatible with
2 M$_{solar}$ neutron stars. Determination of moment of inertia of a neutron 
star in relativistic neutron star binaries in the SKA era would allow the
simultaneous measurements of mass and radius of a particular neutron star. The
model independent construction of an EoS might be possible if masses and radii
of same neutron stars are known \cite{lind92}. This is one of several spin offs
of the knowledge of moment of inertia. The superfluid phase in pulsar glitches 
is another interesting area of investigation. The entrainment effect in the
superfluid matter could severely constrain the reservoir of superfluid moment
of inertia in the crust \cite{ander12}. The recent discovery of negative 
effective mass in a Bose-Einstein condensate makes this study more interesting
and challenging \cite{prl118}. It is to be seen what is the role of negative 
effective mass on the superfluid hydrodynamics in neutron stars and its 
connection to glitch phenomena.        

\section*{Acknowledgement}
The author acknowledges discussions with R. Nandi and P. Char.

%
%

\end{document}